\documentclass[12pt]{article}
\usepackage{amsmath}
\usepackage{latexsym}
\usepackage{amsfonts}
\usepackage[normalem]{ulem}
\usepackage{soul}
\usepackage{array}
\usepackage{amssymb}
\usepackage{extarrows}
\usepackage{graphicx}
\usepackage[backend=biber,
style=numeric,
sorting=none,
isbn=false,
doi=false,
url=false,
]{biblatex}

\usepackage{subfig}
\usepackage{wrapfig}
\usepackage{wasysym}
\usepackage{enumitem}
\usepackage{adjustbox}
\usepackage{ragged2e}
\usepackage[svgnames,table]{xcolor}
\usepackage{tikz}
\usepackage{longtable}
\usepackage{changepage}
\usepackage{setspace}
\usepackage{hhline}
\usepackage{multicol}
\usepackage{tabto}
\usepackage{float}
\usepackage{multirow}
\usepackage{makecell}
\usepackage{fancyhdr}
\usepackage[toc,page]{appendix}
\usepackage[hidelinks]{hyperref}
\usetikzlibrary{shapes.symbols,shapes.geometric,shadows,arrows.meta}
\tikzset{>={Latex[width=1.5mm,length=2mm]}}
\usepackage{flowchart}\usepackage[paperheight=11.0in,paperwidth=8.5in,left=1.0in,right=1.0in,top=1.0in,bottom=1.0in,headheight=1in]{geometry}
\usepackage[utf8]{inputenc}
\usepackage[T1]{fontenc}
\TabPositions{0.5in,1.0in,1.5in,2.0in,2.5in,3.0in,3.5in,4.0in,4.5in,5.0in,5.5in,6.0in,}

\renewcommand{\_}{\kern-1.5pt\textunderscore\kern-1.5pt}

\setlistdepth{9}
\renewlist{enumerate}{enumerate}{9}
		\setlist[enumerate,1]{label=\arabic*)}
		\setlist[enumerate,2]{label=\alph*)}
		\setlist[enumerate,3]{label=(\roman*)}
		\setlist[enumerate,4]{label=(\arabic*)}
		\setlist[enumerate,5]{label=(\Alph*)}
		\setlist[enumerate,6]{label=(\Roman*)}
		\setlist[enumerate,7]{label=\arabic*}
		\setlist[enumerate,8]{label=\alph*}
		\setlist[enumerate,9]{label=\roman*}

\renewlist{itemize}{itemize}{9}
		\setlist[itemize]{label=$\cdot$}
		\setlist[itemize,1]{label=\textbullet}
		\setlist[itemize,2]{label=$\circ$}
		\setlist[itemize,3]{label=$\ast$}
		\setlist[itemize,4]{label=$\dagger$}
		\setlist[itemize,5]{label=$\triangleright$}
		\setlist[itemize,6]{label=$\bigstar$}
		\setlist[itemize,7]{label=$\blacklozenge$}
		\setlist[itemize,8]{label=$\prime$}

\begin{document}
\begin{Center}
\textbf{A Short Commentary}
\end{Center}\par

\begin{Center}
Does purely physical information have meaning? A comment on Carlo Rovelli’s paper $``$Meaning=information + evolution$"$ 
\end{Center}\par

\begin{Center}
Roman Krzanowski
\end{Center}\par

\begin{Center}
Pontifical University of John Paul II, Cracow, Poland
\end{Center}\par

\begin{adjustwidth}{0.81in}{0.5in}
\begin{justify}
The note discusses the concept of meaningful, physical information presented by Carlo Rovelli. It points out certain consequences of the information model not elucidated in the original paper but important to its comprehensive understanding.
\end{justify}\par

\end{adjustwidth}

\vspace{\baselineskip}

\vspace{\baselineskip}
\begin{multicols}{2}
\begin{justify}
Information is mostly perceived as an abstract idea that is closely associated with knowledge, cognition, or communication. It has meaning, intentionality, and purpose, and its presence depends upon the existence of a cognitive agent. Information may also be regarded as a concrete, physical phenomenon. Such information does not need an agent to exist, but it has no meaning, purpose, or intentionality. So how can information be, at the same time, abstract and endowed with meaning while also being concrete and physical yet meaningless? Paul Davies [1] identifies this abstract–concrete dichotomy as a fundamental problem of information. Carlo Rovelli [2], meanwhile, also highlighted this gap between information as a physical phenomenon and information with meaning. In his paper, Rovelli [2] outlines how to close this gap. This commentary examines Rovelli’s claims and asks whether his concept of information offers a fundamental solution to the abstract–concrete dichotomy, the concept of information in general, and the concept of meaningful information. Note that this discussion is intentionally brief, so several ideas are only conveyed through references and not elaborated in detail.
\end{justify}\par

\begin{justify}
Rovelli [2] presents a $``$purely physical definition of meaningful information,$"$ but how? First, Rovelli uses Shannon’s theory of communication (TOC) concept for measuring information and derives from this the $``$relative information$"$  between two physical systems. He denotes it as $``$a purely physical version of the notion of information$"$  or a purely $``$physical correlation.$"$ But what is Rovelli’s pure correlation? Pure correlation is defined as the correlation of two probability distributions (in compatible state spaces) or as the difference between (information) entropies of two probability distributions. As Rovelli states, $``$[these] correlations can exist because of physical laws or because of specific physical situations, or arrangements or mechanisms$ \ldots $ $"$  Therefore, a $``$purely physical correlation,$"$  which is information for Rovelli, represents some (co-)dependence of the (two) states of physical systems. In other words, the two states are correlated, and the entropy difference or joint probability are formal expressions of this relation. 
\end{justify}\par

\begin{justify}
Second, Rovelli proposes (following the Wolpert and Kolchinsky paper [3]) that Darwinian evolution provides the essential mechanisms to endow the natural/physical processes related to the evolution of organisms with some notion of meaning, purpose, and intentionality. From this viewpoint, environmental/physical stimulus has meaning for an organism if it increases its chances of its survival (or preservation). As a very rudimentary example, if bacteria are able to avoid a harmful environment by sensing it through some sort of physical stimuli, interpreting it correctly, and initiating evasive action, it creates meaning out of the physical sensory input. In other words, it creates meaningful information for itself. This is how environmental stimuli acquire meaning, intentionality, and purpose. (In the original text [3], the meaning of information is tied with an organism’s ability to stay outside the $``$thermal equilibrium,$"$  that is, to keep a low entropy state). By combining a Darwinian-based interpretation of meaning, purpose, and intentionality with information perceived as a correlation between physical states using relative information [4], Rovelli defines purely physical meaningful information. In his own words, he creates $``$the crucial first link of the chain$"$  for connecting physical information with meaning, as expressed in the paper’s title: $``$meaning=information + evolution$"$ .
\end{justify}\par

\begin{justify}
So, what is Rovelli’s $``$purely physical definition of meaningful information$"$ ? Rovelli used Shannon’s information measure — Shannon never defined information but rather a measure of information and information entropy—to define the relative information (the information element of Rovelli’s definition) and then applied this to the physical aspect of evolutionary processes, thus obtaining the Darwinian element responsible for meaning in Rovelli’s definition. However, in his definition, Rovelli implicitly adopts the whole framework of Shannon’s communication model (TOC) and the formula for the amount of information in messages between agents (i.e., information entropy). Why is this important? First, the model of relative information derived from Shannon’s TOC information entropy implicitly includes at its basis the concept of the communication model. This implies that (physical) information is created by agents or exchanged between them, so the communication context is essential for the creation of meaning.
\end{justify}\par

\begin{justify}
Another consequence implied in Shannon’s model of information entropy is the probabilistic nature attributed to information. This serves a purpose in the optimization of communication channels, or it may be acceptable in specific cases that deal with very large state spaces, such as the entropy of a black hole [10]. We can ask, however, is information in its very essence really probabilistic, or is probability introduced to characterize how information is really an expression of the epistemic opaqueness of the phenomena (i.e., the case under study) rather than a feature of the information itself?
\end{justify}\par

\begin{justify}
There is also a question as to whether we can attribute meaning, intentionality, and purpose to probabilistic evolutionary processes. What do these terms mean in this context? What is more, do we not indirectly invoke the notion of intelligent design or God’s hand in creation by assigning meaning, intentionality, and purpose (even in a reduced form) to natural processes? Of course, we may propose highly reductionist interpretations of these terms, and this is indeed what Rovelli did, but in this way, these terms would not mean what they purport to denote and what they would then denote in reality is unclear.
\end{justify}\par

\begin{justify}
Why is such a (over) use of the terms’ meaning, intentionality, and purpose not recommended? This is because when claims of meaning, even when qualified, are associated with physical or biochemical processes, it distorts the meaning of meaning. These claims reinterpret the very concept of meaning, which in turn may result, as is often the case, in misinterpretations or misuse. There is also always a danger of running the Sokal affair in reverse. As a reminder, the Sokal affair involved using terms from one domain (e.g., physics, mathematics) in another, completely different context (e.g., social sciences, philosophy).
\end{justify}\par

\begin{justify}
We can observe that the author sensed he was on rather unstable grounds when blending the TOC, physical phenomenon, meaning, and intentionality, and he admits that his ideas are merely proposals. Rovelli also points out that this sort of meaning is a very primitive one. The distance between the theoretical models of the cosmos or Bach’s fugues and a bacteria’s reaction to its environment are light years apart, yet the reaction of the bacteria is, at its core, the seed for others. This is conceptual quicksand for sure. Granting meaning, intentionality, and so on to natural biological systems is, it seems an anthropomorphization of nature. Restricting these concepts to people runs the risk of incurring human exceptionalism. The proper approach is probably somewhere in the mythical middle, such as recognizing the exceptional human cognitive ability but viewing it as an emerging function in a higher level biological system. Another approach is to simply bite the bullet and stick to more widely recognized definitions.
\end{justify}\par

\begin{justify}
Summing up Rovelli’s view, information is correlation (pure correlation) between two physical systems (as defined above) and meaningful information is pure correlation in the above sense within the context of an evolutionary process being viewed as a communication process. This approach excludes, in principle, the concept of information where it is an intrinsic, not agent-relative, feature of physical phenomena.
\end{justify}\par

\begin{justify}
Rovelli’s meaningful information is certainly an interesting idea and worthy of study. However, its novelty and import must be placed in perspective. Shannon’s TOC naturally lends itself to the concept of physical information in evolutionary processes, because communication is a physical process at its foundations and the very essence of evolution is based on communication between the environment and living agents. It must be said, however, that Shannon probably did not intend to make an association between purely physical processes, evolution and information [7]. As we said, Darwinian evolution is in some sense just a specific instance of a communication model being adopted for an evolutionary process. It is rather obvious that we are the creations of evolution (and therefore created through communication), and our cognitive faculties are the product of this (see e.g., Nagel [5], Searle [6]). 
\end{justify}\par

\begin{justify}
Now, does this proposal solve the $``$gap$"$  problem? Perhaps it does, but at the cost of stretching the meaning of meaning itself and forcing the concept of physical information into the TOC framework. Is this a price we want to pay, though? It seems that the solution to the $``$gap$"$  lies somewhere else.
\end{justify}\par

\begin{justify}
It has been suggested that the abstract–concrete dichotomy of information is caused by a misconceptualization of what information is [8]. One possible solution to resolve this abstract–concrete tension could be based on recognizing that information at the fundamental level is an organization or other form of physical phenomena [8], and it does not require the communication model framework to be defined, so there is no need for the TOC. This information exists everywhere where there is physical reality, as anything in reality takes some form, so in this view, information is fundamental. In addition, this information is ontologically objective in Searle’s sense [6]—it exists independently of anyone $``$sensing it$"$  but is meaningless. The meaning in, or of, information is created by a cognitive agent, for an agent, or in the agent’s cognitive system. From this perspective, the dichotomy dissolves, because information is fundamentally physical, but meaning is created by an agent and stays with him or her.
\end{justify}\par

\begin{justify}
Assuming that meaning and intentionality are properties created, or attributed, by cognitive capacities, the presence of an agent with cognitive ability endows physical information with meaning. In other words, the agent derives meaning from neutral physical stimuli. Meaningful information is them simply agent-relative or ontologically subjective [6], even if it is shared among many agents. How meaning is created by cognitive agents is not precisely known at this point, but we may assume that meaning has a biological basis, because cognition is a biological phenomenon. We need to avoid dualism, however, because this would take us into Never-never land of Descartes.
\end{justify}\par

\begin{justify}
Now, does this agent-originated meaning correlate with physical information? Yes, it does, because cognitive natural/biological agents evolve to respond to the external environment by interpreting it as something meaningful for them. However, can we assign meaning to any response by a biological agent to environmental stimuli like in [2]? If we were to, we would run the risk of trivializing the concept of meaning, because we would assign meaning, knowledge, and intelligence (or some derivatives thereof) to plants, single-cell organisms, and so on, which is a trend in biology that is actually happening (see e.g., [9]). Can we really say that meaning is a well-defined concept that spans the continuum from the primitive responses of simple biological systems to complex human-based meaning? We could say so, and some are indeed doing precisely this [9], but it is tantamount to proposing some form of panpsychism. 
\end{justify}\par

\begin{justify}
In summary, we can see how introducing the concept of relative information based on the TOC and associating with it a meaning of sorts created through the evolutionary process does dissolve the alleged gap, albeit only superficially. Indeed, it prompts more questions about meaning and information, particularly physical information, than it resolves. Introducing this new concept of meaning does not necessarily help to explain what comprises meaningful information and what meaning is, and defining physical information in the context of the TOC renders physical information relative rather than absolute (ontologically objective), which is what physical phenomena like information are.
\end{justify}\par

\begin{justify}
Acknowledgements
\end{justify}\par

\begin{justify}
I would like to thank Prof. Pawel Polak and Dr. Jacob Krzanowski for their critical reading of this commentary and their very helpful suggestions. Any mistakes or inaccuracies are mine alone.
\end{justify}\par

\vspace{\baselineskip}

\end{multicols}
\begin{justify}
References
\end{justify}\par

[1] P. Davies. 2019. \textit{The Demon in the machine}. Allen Lane, New York.\par

\begin{justify}
[2] C. Rovelli. 2016. Meaning = information + evolution. arXiv:1611.02420v1. [On line]Available at \href{https://arxiv.org/pdf/%201611.02420.pdf}{https://arxiv.org/pdf/ 1611.02420.pdf}, accessed on 3.28.2020.
\end{justify}\par

\begin{justify}
[3] D. H. Wolpertand A. Kolchinsky, A. 2016. Observers as systems that acquire information to stay out of equilibrium, in$``$The physics of the observer$"$  Conference. Banff, August 17- 22 2016. [On line] Available at  watch?v=zVpSAjAe-tE, accessed on 01.04.2020.
\end{justify}\par

\begin{justify}
[4] C. E. Shannon. 1948. \textit{A Mathematical Theory of Communication}.The Bell System Technical Journal.Vol.27.S.379–423., 1948.
\end{justify}\par

\begin{justify}
 [5] T. Nagel. 2012. Mind and Cosmos: Why the materialist neo-Darwinian conception of nature is almost certainly false. Oxford New York: Oxford University Press
\end{justify}\par

{\fontsize{11pt}{13.2pt}\selectfont [6] J. Searle. 1998. \textit{Mind, Language and Society}. Basic Books, New York.\par}\par

{\fontsize{11pt}{13.2pt}\selectfont [7] C. E. Shannon. 1956. The Bandwagon. IRETransactions – Information Theory.p.3.\par}\par

{\fontsize{11pt}{13.2pt}\selectfont [8] R. Krzanowski. 2020. Why can information not be defined as being purely epistemic? To be published.\par}\par

{\fontsize{11pt}{13.2pt}\selectfont [9] A. Trewavas.2017. The foundations of plant intelligence. Interface Focus 7: 20160098. \href{http://dx.doi.org/10.1098/rsfs.2016.0098}{http://dx.doi.org/10.1098/rsfs.2016.0098}\par}\par

{\fontsize{11pt}{13.2pt}\selectfont [10] \par}J. D. {\fontsize{11pt}{13.2pt}\selectfont Barrow. 2007. New Theories of everything. Oxford University Press, Oxford.\par}\par

\vspace{\baselineskip}

\printbibliography
\end{document}